\documentclass[cameraready]{Interspeech}

\usepackage{booktabs}
\usepackage{multirow}
\usepackage{graphicx}
\usepackage{float}
\usepackage{amssymb}
\usepackage{cite}
\usepackage{siunitx}
\hypersetup{colorlinks=true,linkcolor=blue,citecolor=blue,urlcolor=blue}
\title{CLAR: CIF-Localized Alignment for Retrieval-Augmented Speech LLM-Based Contextual ASR}

\author[affiliation={1}]{Shangkun}{Huang}
\author[affiliation={1}]{Huan}{Shen}
\author[affiliation={1}, correspondingauthor]{Wei}{Zou}
\author[affiliation={1}]{Yunzhang}{Chen}

\address{
  $^1$ BRVoice Team, Bairong, Inc., China
}

\email{shangkun.huang@brgroup.com, wei.zou@brgroup.com}

\keywords{Contextual ASR, Retrieval-Augmented Speech Recognition, Hotword Retrieval, CIF, Weakly Supervised Alignment}

\usepackage{comment}

\usepackage{CJKutf8}
\AtBeginDocument{
\begin{CJK*}{UTF8}{gbsn}}
  \AtEndDocument{
\end{CJK*}}

\begin{document}

\maketitle

\begin{abstract}

  \noindent Speech LLM-based ASR often struggles with named entities and long-tail words due to strong internal language-model priors. Retrieval-augmented biasing can help, but its effectiveness depends on accurate hotword localization in full-utterance speech under weak supervision. We propose CLAR, a dual-encoder speech-text retriever that uses Continuous Integrate-and-Fire (CIF) to learn monotonic token-level alignments without timestamps. With length-aware localized matching, CLAR anchors short-entity acoustic cues and reduces representation dilution and attention drift. The retriever is trained with a multi-granularity objective combining global and local segment-level contrastive losses and a CIF quantity constraint. At inference, top-ranked hotwords are injected as contextual prompts for the Speech LLM, improving recognition without shallow fusion. Experiments show that CLAR significantly improves hotword retrieval and reduces both CER and B-WER against strong contextual ASR baselines.

\end{abstract}


\section{Introduction}
\label{sec:introduction}

Generative large language models (LLMs) have driven remarkable progress in general-domain ASR \cite{huang2025leveraging,cfl,br-asr,gong2024contextual,xu2025fireredasr,jia2025efficient-liullm,shi2026qwen3-asr,an2025funasr}. However, in open-domain settings, accurate recognition of named entities and long-tail words remains a formidable challenge. These terms are typically low-frequency, sparsely distributed, and lack rich semantic context. As a result, strong internal language-model priors often dominate end-to-end decoding, overwhelm weak acoustic evidence, and bias outputs toward high-frequency common words. This ``semantic over-reliance'' can induce entity substitutions and hallucinations, substantially undermining the robustness and reliability of downstream applications such as information retrieval, question answering, and conversational interaction.

To mitigate recognition failures caused by overly strong semantic priors, prior work frames Speech-LLM decoding as prompt-conditioned generation with contextual biasing or retrieval-augmented generation (RAG). Injecting hotword candidates or few-shot examples can improve access to low-frequency words \cite{sun2024contextual-1}, but performance remains sensitive to candidate-list precision and may degrade when the list grows \cite{kong2025glclap,gong2024contextual}. Representative solutions include CTC-based auxiliary filtering, local context correction, and multi-domain modeling \cite{cfl,trinh2025improving-5,bai2024seed}, as well as RAG pipelines using textual context mining, entity detection, vector retrieval, or phoneme-level edit-distance matching \cite{siskos2025retrieval,lei2025contextualization,pusateri2025retrieval,an2026retrieval}.

\begin{figure}[t]
  \centering
  \includegraphics[width=0.92\linewidth]{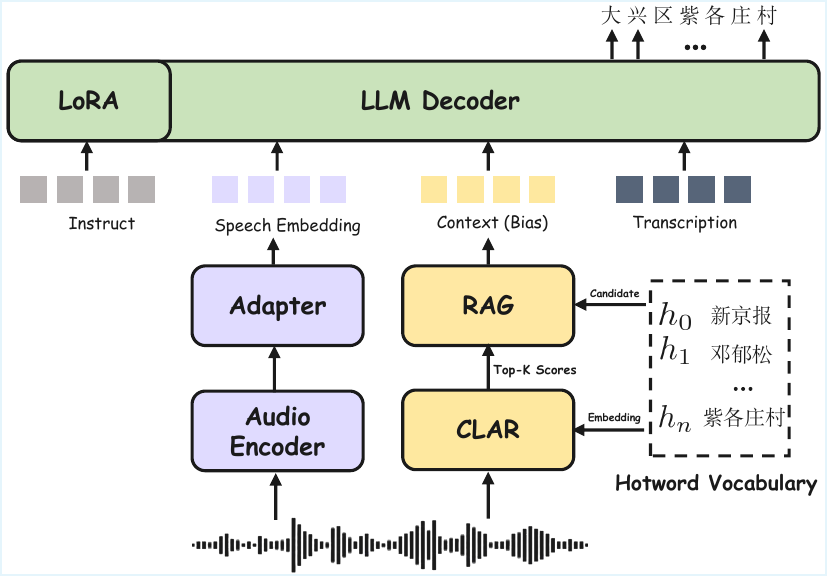}

  \caption{CLAR retrieval-augmented contextual ASR: CIF-Localized Alignment retriever for localized hotword matching and prompt-based decoding with a Speech LLM.}
  \vspace{-0.5em}
  \label{fig:overall_architecture}
  \vspace{-1.2em}
\end{figure}

Ultimately, the efficacy of these contextual and retrieval-augmented methods hinges on the front-end retriever's ability to stably and accurately localize hotword segments within full-utterance speech. Under weak supervision (without timestamps), existing approaches still face two critical bottlenecks. First, a mismatch in representation granularity causes feature ``dilution'': conventional dual-tower retrievers apply global pooling, compressing variable-length speech into a single utterance-level vector. In full-utterance speech, the brief acoustic cues of short entities (often lasting only a few hundred milliseconds) are easily averaged out by background noise or irrelevant content, and this issue is further aggravated when speech--text mapping is learned primarily from TTS-generated data \cite{br-asr,luo2025generative}. This mismatch can further weaken robustness under real-world acoustic scenarios. Second, weak supervision induces alignment deviation. Lacking explicit temporal constraints, unconstrained soft-matching paradigms capture global semantic co-occurrence rather than strict temporal correspondence, causing focus drift and misaligned local aggregation.

As shown in \autoref{fig:overall_architecture}, we propose CLAR, a dual-encoder speech--text retriever for localized hotword matching, whose top candidates are injected as prompts into a Speech LLM for contextual ASR; the details are provided in \autoref{sec:method}. The core technical insight of CLAR is a CIF-Localized Alignment module that learns token-level monotonic boundaries without timestamps. This enables length-aware localized window matching that precisely anchors short-entity cues, substantially reducing representation dilution and attention drift, as illustrated in \autoref{fig:clap_architecture}. To decouple global semantics from fine-grained details while reducing reliance on synthetic data, we optimize the retriever on real ASR speech using a multi-granularity objective that combines global and fragment-level contrastive losses with a CIF quantity constraint.

Our main contributions are summarized as follows:
\begin{itemize}
  \item We propose CLAR, a CIF-localized dual-encoder retriever that learns monotonic token-level alignments and performs length-aware local matching under weak supervision, while remaining plug-and-play across Speech LLM backbones.
  \item We design a multi-granularity training objective that couples global contrastive loss, segment-level contrastive loss, and a CIF quantity constraint to improve fine-grained retrieval accuracy.
  \item We show that CLAR delivers consistent gains in both hotword retrieval and contextual ASR, achieving state-of-the-art contextual ASR results over strong baselines.
\end{itemize}

\section{Proposed Method}


\subsection{Overall Framework}
\label{sec:method}
\begin{figure}[t]
  \centering
  \includegraphics[width=0.92\linewidth]{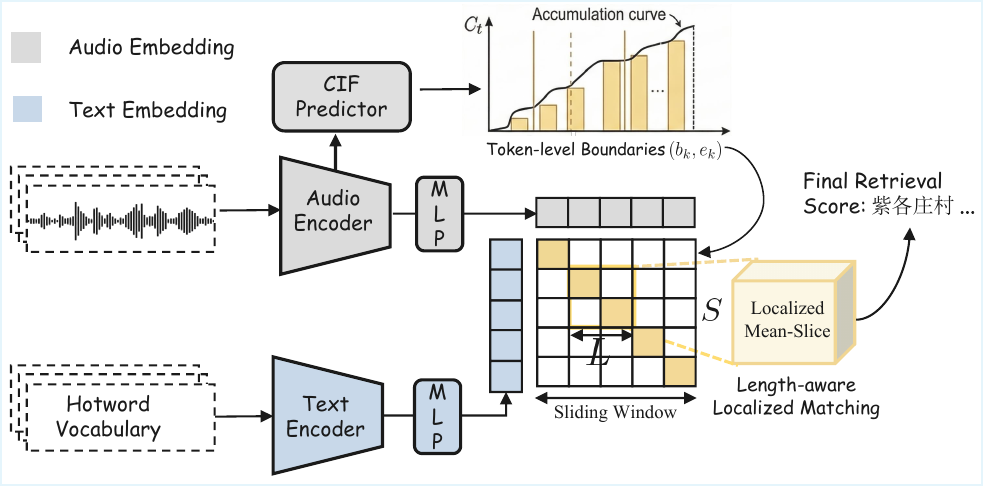}
  \caption{The CLAR retriever with dual encoders, CIF-based monotonic alignment, and length-aware localized matching for hotword retrieval.}
  \label{fig:clap_architecture}
  \vspace{-1.2em}
\end{figure}

\autoref{fig:overall_architecture} shows the retrieval-augmented pipeline, where a CIF-Localized Alignment CLAR module retrieves hotword candidates and a Speech LLM decodes the final transcript with prompt-based biasing.
\autoref{fig:clap_architecture} details the retriever, including the dual-encoder backbone, CIF-based monotonic alignment, and localized matching for fine-grained hotword retrieval.
The framework consists of two components: (1) a dual-encoder CLAP-style retriever with a CIF alignment mechanism to provide token-level temporal boundaries; and (2) a general-purpose Speech LLM, which consumes the retrieved candidates as contextual prompts to improve entity recognition.
We next describe the retriever and alignment mechanism, then detail the localized matching objective and training strategy; the full retrieval procedure is summarized in Algorithm~1.

\subsection{CLAP-Style Dual-Encoder and CIF-Localized Alignment}
\label{sec:clap_cif}

\textbf{Dual-encoder architecture.} The CLAR module follows the CLAP\cite{clap} framework and adopts a dual-tower design consisting of a speech encoder and a text encoder. The speech encoder is instantiated as a Paraformer\cite{gao2022paraformer} encoder based on the Self-Attention with Non-autoregressive Modeling architecture. It takes log-mel filterbank features $\mathbf{X} \in \mathbb{R}^{T_0 \times F}$ as input and produces frame-level hidden representations $\mathbf{H} = [\mathbf{h}_1, \dots, \mathbf{h}_T] \in \mathbb{R}^{T \times D_{\text{enc}}}$, where $T$ denotes the downsampled frame count and $D_{\text{enc}}$ is the encoder hidden dimension. The text encoder is initialized from Chinese-RoBERTa-wwm-ext, a pretrained RoBERTa-based Transformer encoder that maps a token sequence to contextualized representations, and a sentence-level embedding is obtained via masked mean pooling over valid token positions.

\begin{table}[t]
  \label{tab:cif_clap_pseudo}
  \centering
  \footnotesize
  \setlength{\tabcolsep}{3pt}
  \renewcommand{\arraystretch}{1.02}
  \begin{tabular}{p{0.97\linewidth}}
    \toprule
    \textbf{Algorithm 1: CIF-Localized Alignment CLAR for hotword retrieval} \\
    \midrule
    \textbf{Input:} speech features $\mathbf{X}$, hotword candidates $\{\mathbf{y}_j\}_{j=1}^{M}$, candidate token lengths $L_j$, and top-$K$. \\
    \textbf{Output:} ranked hotword list $\mathcal{C}_K$. \\
    \textbf{1.} Encode speech and predict CIF alignment: $\mathbf{H} \leftarrow \text{SpeechEnc}(\mathbf{X})$, $\boldsymbol{\alpha}\leftarrow\text{CIFPred}(\mathbf{H})$, and obtain token boundaries $\{(b_s,e_s)\}_{s=1}^{S}$. \\
    \textbf{2.} Project and normalize embeddings: $\mathbf{a}_t \leftarrow \text{Norm}(f_a(\mathbf{h}_t))$ and $\mathbf{e}_j \leftarrow \text{Norm}(f_t(\text{TextEnc}(\mathbf{y}_j)))$. \\
    \textbf{3.} Compute frame--candidate similarity: $S_{t,j} \leftarrow \tau\,\mathbf{a}_t^\top \mathbf{e}_j$. \\
    \textbf{4.} For each candidate $j$, perform length-aware localized scoring: $\text{score}_j(s)=\frac{1}{e_{s+L_j-1}-b_s+1}\sum_{t=b_s}^{e_{s+L_j-1}}S_{t,j}$, and set $\hat r_j \leftarrow \max_s \text{score}_j(s)$. \\
    \textbf{5.} Rank by $\hat r_j$ and return $\mathcal{C}_K \leftarrow \text{TopK}(\{\hat r_j\},K)$. \\
    \bottomrule
  \end{tabular}
  \renewcommand{\arraystretch}{1.0}
  \vspace{-1.2em}
\end{table}

Both modalities are mapped into a shared embedding space of dimension $D$ through dedicated projection layers. Specifically, the audio projection $f_a(\cdot)$ and the text projection $f_t(\cdot)$ are each realized as a two-layer feed-forward network with a ReLU activation:
\begin{equation}
  \mathbf{a}_t = \frac{f_a(\mathbf{h}_t)}{\lVert f_a(\mathbf{h}_t)\rVert_2}, \quad
  \mathbf{e}_j = \frac{f_t(\mathbf{z}_j)}{\lVert f_t(\mathbf{z}_j)\rVert_2},
\end{equation}

where $\mathbf{z}_j$ denotes the pooled output of the text encoder for the $j$-th candidate, and the resulting embeddings $\mathbf{a}_t, \mathbf{e}_j \in \mathbb{R}^{D}$ are $\ell_2$-normalized.

\textbf{CIF-based monotonic alignment}. A central component of our system is the CIF predictor~\cite{dong2020cif}, which operates on encoder outputs $\mathbf{H}$ to produce frame-level weights $\boldsymbol{\alpha}=[\alpha_1,\dots,\alpha_T]$, where $\alpha_t\in(0,1)$. The weights are generated by a 1D convolution, followed by a linear layer and sigmoid activation:
\begin{equation}
  \alpha_t = \sigma\left(\text{Linear}\left(\text{ReLU}\left(\text{LayerNorm}\left(\text{Conv1d}(\mathbf{H})\right)\right)\right)_t\right).
\end{equation}
CIF accumulates these weights through a counter $C_t=C_{t-1}+\alpha_t$. A \textit{fire} event is triggered when $C_t\geq\theta$ (default $\theta=1.0$), which defines a monotonic acoustic boundary; the residual is carried over by $C_t\leftarrow C_t-\theta$. For structured processing, frame $t$ can be mapped to a token index by counting prior fire events,
\begin{equation}
  \text{tok}(t)=1+\sum_{j=1}^{t-1}\mathbf{1}[C_j\geq\theta],
\end{equation}
which yields a non-decreasing frame-to-token mapping and supports length-aware local windows for downstream speech--text matching.

During training, the CIF predictor uses a scaling strategy in which the predicted weight sequence $\boldsymbol{\alpha}$ is multiplied by a scalar so that the total accumulated weight $\sum_t \alpha_t$ matches the ground-truth token length; this teacher-forcing constraint stabilizes alignment learning. The CIF-related parameters are initialized from the corresponding Paraformer component.


\subsection{CIF-Localized Hotword Retrieval}
\label{sec:cif_localized}
This subsection presents the core contribution of our method: a CIF-localized hotword retrieval mechanism that leverages token-level alignment for fine-grained, variable-length cross-modal matching.

\textbf{Cross-modal similarity matrix}. Given a speech utterance with projected, normalized frame-level embeddings $\mathbf{A} = [\mathbf{a}_1, \dots, \mathbf{a}_T] \in \mathbb{R}^{T \times D}$ and a hotword vocabulary bank $\mathbf{E} = [\mathbf{e}_1, \dots, \mathbf{e}_N] \in \mathbb{R}^{N \times D}$ containing $N$ precomputed text embeddings, the frame-level cross-modal similarity matrix is computed as:

\begin{equation}
  \mathbf{S} = \tau \cdot \mathbf{A}\mathbf{E}^\top \in \mathbb{R}^{T \times N},
\end{equation}
where $\tau$ is a learnable logit-scale parameter (the inverse temperature) that controls the sharpness of the similarity distribution.

\textbf{Token-span mean-slice aggregation}. The key innovation lies in how the similarity matrix $\mathbf{S}$ is aggregated into a single retrieval score per candidate hotword. Unlike conventional global pooling, which can dilute fine-grained local semantics, our method uses CIF-predicted token boundaries for localized sliding-window aggregation.

CIF alignment maps each token index $k$ to an acoustic frame span $[b_k, e_k]$, where $b_k$ and $e_k$ are the first and last frames aligned to token $k$, respectively. The CIF predictor is configured to emit acoustic tokens at the same granularity as the text encoder tokenizer. For a candidate hotword with text-token length $L$, we slide an acoustic window of width $L$ over the CIF-emitted token sequence. At each window position $s \in \{0, 1, \dots, K-L\}$, where $K$ is the total number of CIF-emitted tokens, the corresponding acoustic span is $[b_s, e_{s+L-1}]$. The mean-slice score for candidate $j$ at window position $s$ is defined as:

\begin{equation}
  \text{score}_j(s) =
  \frac{1}{e_{s+L-1}-b_s+1}\sum_{t=b_s}^{e_{s+L-1}} S_{t,j}.
\end{equation}
The final retrieval score for the $j$-th candidate is obtained by taking the maximum over all valid window positions:
\begin{equation}
  \hat{r}_j = \max_{s}\text{score}_j(s).
\end{equation}
This formulation has two advantages. First, it supports variable-length matching by aligning each candidate to a length-proportional window. Second, localized aggregation preserves salient local cues and avoids global-pooling dilution, while using prefix-sum window scoring with length-wise candidate batching for practical computation.

\subsection{Training Objectives and Retrieval-Augmented Decoding}
\label{sec:training_decoding}
\textbf{Training objectives}. The CLAR retriever is trained with a multi-level contrastive learning framework. The total training loss is a weighted combination of three terms:
\begin{equation}
  \mathcal{L} =
  \mathcal{L}_{\text{local}} +
  \mathcal{L}_{\text{global}} +
  \mathcal{L}_{\text{cif}}.
\end{equation}

The local hotword-level contrastive loss $\mathcal{L}_{\text{local}}$ provides fine-grained supervision at the hotword-span level. For each sample, CIF-derived token alignment identifies the annotated hotword frames, which are mean-pooled into a hotword-level speech embedding and contrasted with the corresponding hotword text embedding using a symmetric cross-entropy objective:
\begin{equation}
  \mathcal{L}_{\text{local}} = \frac{1}{2}\left(\mathcal{L}_{\text{a}\to\text{t}} + \mathcal{L}_{\text{t}\to\text{a}}\right).
\end{equation}
Here, $\mathcal{L}_{\text{a}\to\text{t}} = -\frac{1}{B}\sum_{i}\sum_{j} y_{ij}\log\frac{\exp(\tau\,\mathbf{a}_i^\top \mathbf{e}_j)}{\sum_{j'}\exp(\tau\,\mathbf{a}_i^\top \mathbf{e}_{j'})}$, $\mathcal{L}_{\text{t}\to\text{a}}$ is defined symmetrically, and $y_{ij}$ denotes the soft target derived from sample identity.

The global sentence-level contrastive loss $\mathcal{L}_{\text{global}}$ aligns the utterance-level speech embedding (from masked average pooling over valid encoder frames) with the sentence-level text embedding to preserve global semantics. Introducing the global objective incorporates full utterance-level semantic context and further improves local feature discrimination. Meanwhile, the CIF quantity loss $\mathcal{L}_{\text{cif}}$ regularizes alignment by minimizing the L1 distance between the predicted token count $\sum_t \alpha_t$ and the ground-truth text length, stabilizing CIF boundaries during training.

\textbf{Retrieval-augmented decoding}. At inference time, the trained CLAR retriever processes input speech to produce frame-level embeddings and CIF alignments. We then apply the localized mean-slice aggregation in \autoref{sec:cif_localized} to score all candidates in the hotword vocabulary. The top-$K$ candidates are concatenated into a text prompt and prepended to the Speech LLM instruction as dynamic contextual bias. After SFT of the base Speech LLM, decoding is performed autoregressively conditioned on both the audio input and the augmented prompt. CLAR is used only as an inference-time biasing strategy, with no further parameter updates, shallow fusion, or lattice rescoring required during decoding. Because contextual information is injected through the natural-language prompt interface, the method remains modular and readily transferable across prompt-conditioned Speech LLMs.

\section{Experiments}

\subsection{Dataset and Metrics}
\label{sec:dataset_metrics}

Several open-source datasets have advanced contextual ASR research\cite{bu2017aishell,tang2021kespeech,du2018aishell2}. We use AISHELL-1\cite{bu2017aishell} to train the Speech LLM and AISHELL-2\cite{du2018aishell2} to train CLAR, excluding training utterances that contain AISHELL-1-NE hotwords. The hotword set is AISHELL-1-NE\cite{seaco}. Test-AISHELL-1-NE contains 808 utterances, 400 hotwords, and 226 R1 hotwords, while Dev-AISHELL-1-NE contains 1334 utterances, 600 hotwords, and 371 R1 hotwords. R1 hotwords are those whose recall under the base ASR model is below 40\%.

Following prior work\cite{fu2025pac,le2021contextualized}, we adopt CER (character error rate), B-WER (biased-word error rate), and U-CER (character error rate on non-bias words) as the core evaluation metrics, measuring overall recognition accuracy, bias-word accuracy, and non-bias-word accuracy, respectively. We additionally report Top-1/Top-5/Top-10 recall and F1 score to evaluate hotword retrieval performance.

\subsection{Experimental Configuration}

We use Step-Audio2-mini \cite{stepaudio} as the base Speech LLM. To simulate hotword-injected inference, we randomly sample 5--10 transcript entries as hotwords, each a random 3--6 character n-gram. The Speech LLM is LoRA\cite{hu2022lora} fine-tuned (r=16, alpha=16, target=all\_linear, dropout=0.1, bias=none) on 4 NVIDIA H20 GPUs (batch size 32, 2000 steps) using AdamW, a peak learning rate of 2e-6, a 0.1 warmup ratio, and BF16 mixed precision. For CLAR, the audio and text encoders are initialized from Paraformer\footnote{\href{https://modelscope.cn/models/iic/speech_paraformer-large_asr_nat-zh-cn-16k-common-vocab8404-pytorch}{ModelScope iic/speech\_paraformer-large}} and Chinese-RoBERTa-wwm-ext\footnote{\href{https://huggingface.co/hfl/chinese-roberta-wwm-ext}{Hugging Face hfl/chinese-roberta-wwm-ext}}. CLAR is trained on 4 NVIDIA H20 GPUs (batch size 256) using AdamW with a learning rate of $5 \times 10^{-4}$. Training has two stages: (1) CIF pretraining for 30 epochs with global and local loss weights fixed to 0.0, optimizing only the CIF loss; and (2) 60-epoch joint training initialized from stage (1) to optimize the audio encoder, text encoder, and CIF module.

\subsection{Results and Discussion}
\subsubsection{Hotword Retrieval Performance}

\vspace{-0.6em}

\begin{table}[H]
  \caption{Ablation of loss components for hotword retrieval recall (\%) on AISHELL-1-NE.}
  \vspace{-0.5em}
  \label{tab:ablation_loss}
  \centering
  \footnotesize
  \setlength{\tabcolsep}{4pt}
  \renewcommand{\arraystretch}{1.06}
  \begin{tabular}{ccc ccc ccc}
    \toprule
    \multicolumn{3}{c}{\textbf{Loss Terms}} &
    \multicolumn{3}{c}{\textbf{Test-AISHELL-1-NE}} &
    \multicolumn{3}{c}{\textbf{Dev-AISHELL-1-NE}} \\
    \cmidrule(lr){1-3} \cmidrule(lr){4-6} \cmidrule(lr){7-9}
    $\mathcal{L}_{\text{global}}$ & $\mathcal{L}_{\text{local}}$ & $\mathcal{L}_{\text{cif}}$ &
    R@1 & R@5 & R@10 &
    R@1 & R@5 & R@10 \\
    \midrule
    \checkmark & $\times$ & $\times$ & 33.29 & 54.70 & 62.62 & 29.99 & 47.53 & 58.85 \\
    $\times$ & \checkmark & \checkmark & 96.29 & 98.89 & 99.38 & 95.43 & 98.88 & 99.25 \\
    \checkmark & \checkmark & \checkmark & \textbf{97.03} & \textbf{99.75} & \textbf{99.75} & \textbf{95.73} & \textbf{98.88} & \textbf{99.48} \\
    \bottomrule
  \end{tabular}
  \renewcommand{\arraystretch}{1.0}
  \vspace{-0.9em}
\end{table}

To validate CLAR's internal design, \autoref{tab:ablation_loss} presents an ablation study of the loss components. With only the global contrastive loss $\mathcal{L}_{global}$, Recall@1 reaches 33.29\% on Test-AISHELL-1-NE and 29.99\% on Dev-AISHELL-1-NE. Adding the local contrastive loss $\mathcal{L}_{local}$ and the CIF quantity constraint $\mathcal{L}_{cif}$ yields substantial gains, increasing test-set Recall@1 and Recall@5 to 97.03\% and 99.75\%, respectively. These results indicate that the fine-grained monotonic speech--text alignment induced by CIF effectively mitigates attention drift in weakly supervised soft matching, while the global objective incorporates full utterance-level semantic context and further improves local feature discrimination.

\begin{table}[H]
  \caption{Comparison of F1 scores (\%) on the test Test-AISHELL-1-NE dataset.}
  \vspace{-0.5em}
  \label{tab:f1_score}
  \centering
  \footnotesize
  \setlength{\tabcolsep}{2pt}
  \begin{tabular}{@{}lccc@{}}
    \toprule
    \textbf{Dataset} & \textbf{SeACo-Paraformer\cite{seaco}} & \textbf{GLCLAP\cite{kong2025glclap}} & \textbf{CLAR} \\
    \midrule
    Test Aishell-1 NT & 96.00 & 96.96 & \textbf{97.03} \\
    \bottomrule
  \end{tabular}
  \vspace{-0.6em}
\end{table}

\begin{figure}[H]
  \centering
  \vspace{-0.5em}
  \includegraphics[width=0.86\linewidth]{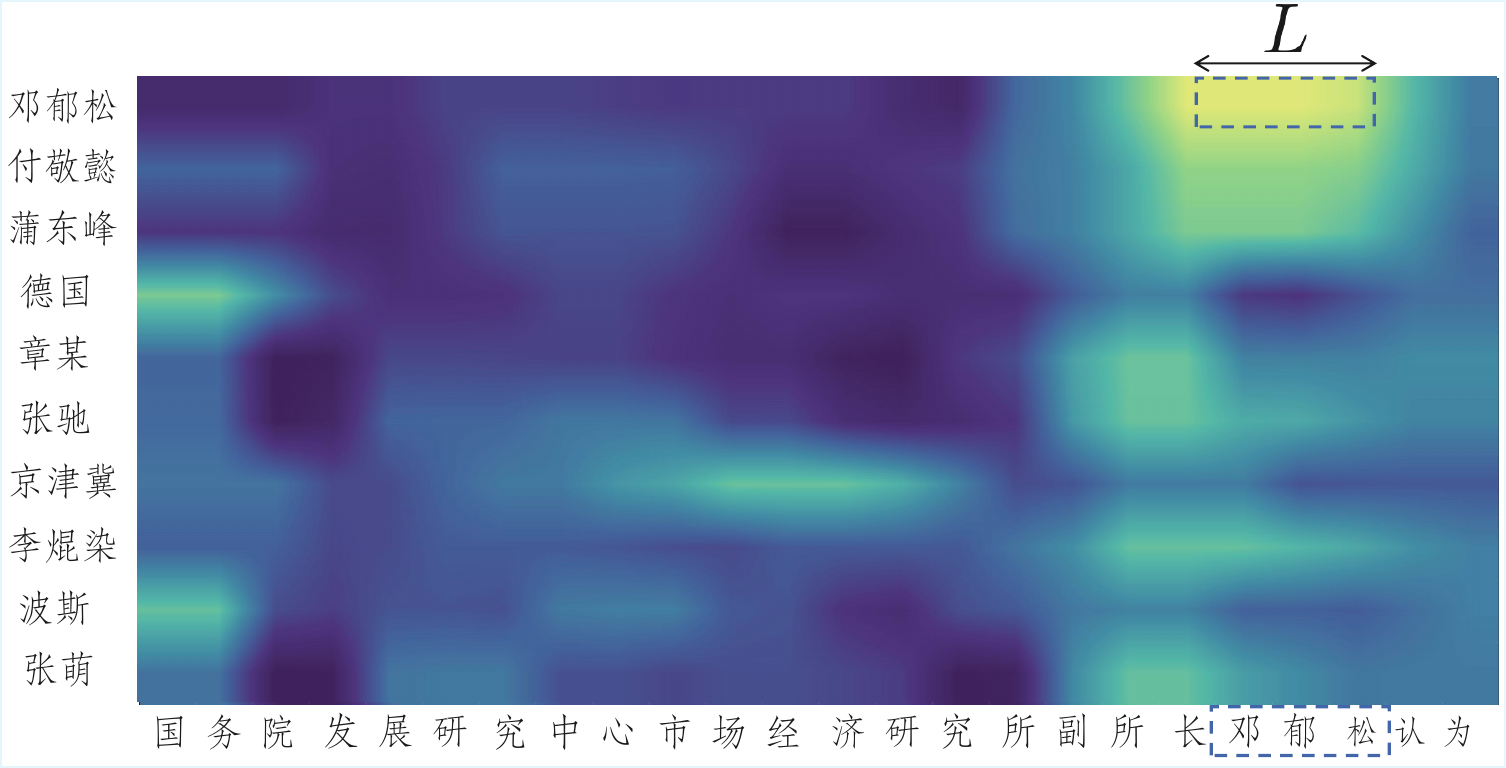}
  \caption{Case-study similarity map for hotword localization.}
  
  \label{fig:fenxi1}
  \vspace{-1.0em}
\end{figure}

To compare retrieval methods, \autoref{tab:f1_score} shows that CLAR achieves the best F1 (97.03\%), slightly surpassing GLCLAP (96.96\%) and outperforming SeACo-Paraformer (96.00\%). Notably, CLAR is trained with roughly one order of magnitude less data than GLCLAP, yet still performs slightly better. \autoref{fig:fenxi1} provides qualitative support: high local speech--text similarity concentrates on the target hotword region (dashed boxes), while non-target candidates remain weak, consistent with improved short-entity localization and reduced attention drift.

\vspace{-0.9em}

\subsubsection{Contextual ASR Results}

\begin{table}[t]
  \caption{Impact of distractor-list size on contextual ASR (\%) on AISHELL-1-NE.}
  \vspace{-0.5em}
  \label{tab:distractors}
  \centering
  \footnotesize
  \setlength{\tabcolsep}{1.5pt}
  \begin{tabular*}{\linewidth}{@{\hspace{0.6em}}p{0.40\linewidth}S[table-format=1.2,table-column-width=0.17\linewidth]S[table-format=2.2,table-column-width=0.19\linewidth]S[table-format=1.2,table-column-width=0.17\linewidth]@{}}
    \toprule
    \textbf{Distractor Setting} & {\textbf{CER} $\downarrow$} & {\textbf{B-WER} $\downarrow$} & {\textbf{U-CER} $\downarrow$} \\
    \midrule
    \multicolumn{4}{@{\hspace{0.6em}}l@{}}{\textbf{Test-AISHELL-1-NE}} \\
    Vanilla Step-Audio2-mini & 1.86 & 12.92 & 0.80\\
    NA            & 1.82 & 13.03 & 0.76 \\
    GT            & 0.82 &  2.34 & 0.78 \\
    $N=226$ (R1)  & 1.36 &  5.97 & 1.03 \\
    $N=400$       & 0.92 &  2.78 & 0.83 \\
    \midrule
    \multicolumn{4}{@{\hspace{0.6em}}l@{}}{\textbf{Dev-AISHELL-1-NE}} \\
    Vanilla Step-Audio2-mini & 1.76 & 13.09 & 0.74 \\
    NA            & 1.80 & 13.09 & 0.75 \\
    GT            & 0.71 &  1.79 & 0.70 \\
    $N=371$ (R1)  & 1.29 &  6.29 & 1.01 \\
    $N=600$       & 0.87 &  3.17 & 0.75 \\
    \bottomrule
  \end{tabular*}
  \vspace{2pt}
  \begin{minipage}{\linewidth}
    \raggedright\footnotesize NA: no bias list; GT: oracle hotword list.
  \end{minipage}
  \vspace{-2.5em}
\end{table}

This section evaluates ASR gains from injecting CLAR-retrieved hotword candidates as prompts. As shown in \autoref{tab:distractors}, we compare Vanilla Step-Audio2-mini (base model), NA (SFT-adapted model without a bias list), and retrieval-based settings with different candidate-pool sizes on Test-AISHELL-1-NE and Dev-AISHELL-1-NE. GT (oracle hotword list) serves as an approximate upper bound (e.g., 0.82\% CER and 2.34\% B-WER on the test set), confirming the importance of accurate contextual candidates. Following \autoref{sec:dataset_metrics}, R1 denotes hotwords with base-ASR recall below 40\% (N=226 test; N=371 dev), representing a harder long-tail regime. With full candidate pools (N=400 test; N=600 dev), CLAR delivers the largest gains over Vanilla and NA: CER drops to 0.92\% (from 1.86\%/1.82\%) on test and to 0.87\% (from 1.76\%/1.80\%) on dev, while B-WER drops to 2.78\% (from 12.92\%/13.03\%) and 3.17\% (from 13.09\%/13.09\%), respectively. U-CER remains low (0.83\% test, 0.75\% dev), indicating little impact on non-bias words. However, on the harder R1 subset, U-CER increases to 1.03\%/1.01\% compared with NA (0.76\%/0.75\%), suggesting that false-positive hotwords can occasionally perturb non-bias-word decoding. CLAR requires only front-end retrieval and natural-language prompting, without Speech LLM updates or shallow fusion, making it a lightweight module for contextual ASR.

\begin{table}[H]
  \caption{Comparison with baseline contextual ASR models (\%) on AISHELL-1-NE.}
  \vspace{-0.5em}
  \label{tab:baseline_comparison}
  \centering
  \footnotesize
  \setlength{\tabcolsep}{1.5pt}
  \begin{tabular*}{\linewidth}{@{\hspace{0.6em}}p{0.59\linewidth}S[table-format=1.2,table-column-width=0.17\linewidth]S[table-format=1.2,table-column-width=0.17\linewidth]@{}}
    \toprule
    \textbf{Model} & {\textbf{CER} $\downarrow$} & {\textbf{B-WER} $\downarrow$} \\
    \midrule
    SeACo-Paraformer\cite{seaco} & 2.20 & 6.71 \\
    CFL\cite{cfl} & 1.51 & 6.03 \\
    PAC\cite{fu2025pac} & 1.13 & 3.07 \\
    \textbf{Ours} & \textbf{0.92} & \textbf{2.78} \\
    \bottomrule
  \end{tabular*}
  \vspace{-1.2em}
\end{table}

For overall recognition performance, \autoref{tab:baseline_comparison} compares different systems. CLAR-augmented contextual ASR achieves 0.92\% CER and 2.78\% B-WER, outperforming SeACo-Paraformer, CFL, and PAC and reaching state-of-the-art performance. These results further confirm that high-precision local retrieval can effectively counteract overly strong language-model priors, reduce named-entity substitutions and hallucinations, and improve recognition accuracy.

\section{Conclusion}

This paper presents CLAR, a CIF-localized dual-encoder retriever for retrieval-augmented contextual ASR with Speech LLMs. CLAR learns token-level monotonic alignments without timestamp supervision and uses length-aware localized matching to anchor short-entity acoustic cues and reduce representation dilution in full-utterance speech. With a multi-granularity objective (global contrastive, local contrastive, and CIF quantity constraints), CLAR substantially improves hotword retrieval and achieves state-of-the-art CER and B-WER on AISHELL-1-NE against strong baselines. Future work will extend CLAR to larger multilingual settings and tighter integration with Speech LLMs.


\section{Generative AI Use Disclosure}
The authors did not use generative AI tools to produce a significant part of the manuscript. If any generative AI tools were used for editing or polishing, all authors remain responsible and accountable for the content, claims, and conclusions.

\bibliographystyle{IEEEtran}
\bibliography{./mybib}

\end{document}